\documentclass[letter]{aa}
\usepackage{graphicx,txfonts,color}
\usepackage{natbib}
\newcommand{\pt}{P$_{3}$}
\newcommand{\cs}{$w$}
\newcommand{\vls}{ReVols}
\newcommand{\rfive}{r$_{500}$}
\newcommand{\lxunit}{erg~s$^{-1}$}
\newcommand{\fxunit}{erg~s$^{-1}$~cm$^{-2}$}
\bibpunct{(}{)}{;}{a}{}{,} 
\setcitestyle{comma}
\begin{document}
\title{Disturbed galaxy clusters are more abundant
\\in an X-ray volume-limited sample}
\author{Gayoung Chon$^1$ and Hans B{\"o}hringer$^1$}
\institute{
$^{1}$Max-Planck-Institut f\"ur extraterrestrische Physik, D-85748 Garching, Germany
}
\date{Received 29 August 2017; Accepted 11 September 2017}
\abstract{
X-ray observations of clusters of galaxies have been used to study 
the large-scale structure of our Universe and to test cosmological models. 
In such studies it is critical to understand the unique survey 
selection function correctly.
In comparison to the cluster detection by the Sunyaev-Zel'dovich effect (SZE), 
it has been shown that X-ray observations preferentially detect 
clusters that have cool cores or are more relaxed as opposed to more disturbed 
or non-cool-core clusters found in SZE surveys.
In this Letter we show that it is not the means of detection, X-rays or SZE,
but the sampling strategy, flux-limited or volume-limited surveying, that
makes the difference.
XMM-Newton observations of the REFLEX clusters in our Volume-Limited Sample 
(ReVols) show that the fraction of disturbed clusters, determined by the third 
moment of the power ratios and by centre shifts, is larger by about
a factor of two than that of relaxed clusters. 
In contrast, two flux-limited cluster samples that can be constructed out of 
\vls{} contain more comparable fractions of disturbed and relaxed clusters, 
which differ by only ten per cent.
We use the ratio of the luminosity measured within \rfive{} to 
that measured in the same aperture without the core region as an indicator 
for a cool core and find that the number of non-cool-core clusters is 
comparable to or larger than that of the cool-core clusters in \vls{}. 
In addition, we show that the X-ray luminosity distributions of the disturbed 
and relaxed clusters are distinctly different, and on average, a displacement of 
60\% in luminosity is required to match two distributions. 
Therefore the larger fraction of relaxed and cool-core clusters reported
in previous X-ray surveys does not result from the X-ray detection per se,
but from the fact that these samples were constructed from flux-limited
surveys.
Our findings also suggest that the Malmquist bias correction used in 
cosmological studies with X-ray galaxy clusters could be improved by
taking the morphological fractions of the galaxy cluster population and
their distinct scaling relations into account.
}
\authorrunning{Chon and B{\"o}hringer}
\titlerunning{X-ray clusters are not cool-core biased in ReVols}
\maketitle
%
%
\section{Introduction}

Clusters of galaxies are highly biased peaks of the underlying dark matter
distribution, allowing us to study the large-scale structure of our Universe.
Because their growth is directly influenced by cosmology, we can also use
them to test cosmological models with a cluster survey, which has a
well-understood selection function. 
Among several detection techniques, X-ray observations of clusters reveal the hot
intra-cluster plasma that fills the cluster potential that is
mostly shaped by dark matter. 
Hence X-ray observations of clusters provide a powerful method to detect and
characterise the cluster population.
Equally powerful is a cluster detection through the Sunyaev-Zel'dovich effect (SZE).
The same plasma is traced in the cluster potential, but with a different dependence
on the gas density. 
Since SZE is almost insensitive to the redshift dimming, unlike X-ray observations, 
which are mostly dominated by Bremsstrahlung emission, an SZE survey selection
is effectively mass-limited rather than flux-limited as in a typical X-ray survey.
Because the strength of the X-ray signal is proportional to the square of the
electron density, an X-ray flux-limited survey has a higher detection 
efficiency for clusters that are brighter at the centre given the same 
mass~\citep{pesce90,markevitch98}.
This effect is responsible for a large part of the scatter in the X-ray
luminosity-mass relation~\citep{pratt09,chon12}, which leads to a significant
Malmquist bias in a flux-limited sample.
This Malmquist bias effect should also lead to an enhancement of the fraction
of relaxed and cool-core clusters in flux-limited samples, which can be
understood as follows.
Most of the clusters in a flux-limited survey are detected near the flux-limit.
A scatter of the X-ray luminosity for a given mass will bring more clusters
at the more luminous end of the $L_X$ distribution into the sample,
which is enriched in cool-core clusters, and it will tend to miss
clusters at the less luminous end of the distribution, where predominantly
disturbed and non-cool cores are located.
Because of this effect, studies have shown that flux-limited X-ray samples have 
a so-called cool-core bias (e.g.~\citet{eckert11,rossetti17,santos17}).
Therefore it is very important to explore the results that are obtained when
this strong Malmquist-type of bias effect can be avoided, for example,
in a survey that is volume-limited.
Hence we devised a program to study this effect with a volume-limited sample 
drawn from the REFLEX cluster survey.
In this Letter we investigate the morphologies of the clusters in this 
sample using a combination of centre shifts and the third moment 
of the power ratios and compare to the properties of the clusters in 
a flux-limited sample.
We also use X-ray luminosity ratios between two different apertures 
to measure the fraction of cool cores and the X-ray luminosity distributions 
of the clusters with different morphologies.

In Sect. 2 we describe our study sample and data reduction. 
The structural analysis and its results are presented in Sect. 3, and 
discussions on the cool-core fraction are found in Sect. 4. 
Section 5 investigates X-ray luminosity distributions of clusters with
different morphological types.
The last section summarises our results and provides a perspective
for future work.

\section{Sample selection and data reduction}

We constructed a volume-limited sample (VLS) from the 
flux-limited REFLEX cluster 
survey~\citep{reflex2-1,reflex2-2} with a redshift limit of $z \le$~0.1 
and a luminosity limit of 5$\times$10$^{43}$~\lxunit{}
in the 0.1-2.4~keV energy band (rest frame).
This sample, which we name ReVols (REFLEX-Volume-limited sample), is one of the
largest VLSs that can be constructed from REFLEX with a relatively high cluster 
density.
We thinned out this sample statistically by selecting only every third cluster 
in the third least luminous bin for an affordable XMM-Newton follow-up. 
In addition, we constructed two flux-limited samples (FLS) within this volume
for a comparison between a VLS and an FLS.
FLS1 was constructed with a flux limit of 10$^{-11}$~\fxunit and FLS2
with 1.3$\times 10^{-11}$~\fxunit.
Technically, an FLS should not be constrained in redshift, hence our numerical 
values in this paper include additional ten clusters that are above the redshift 
limit of ReVols.
More details on the construction of this sample and the catalogue with 
the properties of clusters are being prepared (B\"ohringer et al., in prep.).
We obtained relatively deep XMM-Newton observations as well as data from
the XMM-Newton archive for a total of 93 clusters. 
The exposure of individual observations was designed to collect at least 
4000 photons inside \rfive{} of each cluster for reliable structural 
measurements. 
All archival data satisfy this minimum criterium.

We processed the XMM-Newton data as described in~\cite{chon12}.
For the clusters whose \rfive{} is larger than the field-of-view 
of XMM-Newton, we used ROSAT PSPC observations to model the cluster 
emission and estimated the background in the XMM-Newton data.
We comment on these clusters in the analysis section.

\section{Structural analysis}

To quantitatively determine the degree of substructure, we employed
two common substructure measures: power ratios~\citep{buote95}, 
and centre shifts~\citep{poole06}.
They are well tested for X-ray observations and simulations 
(see e.g.~\citet{rexcess_sub,chon12,mahdavi13,rasia13,chon16})
and have been shown to provide very useful diagnostics.

\subsection{Power ratio calculation}

The power ratio method first introduced by~\cite{buote95} was motivated by
the assumption that the X-ray surface brightness closely traces the projected
two-dimensional mass distribution of a cluster. 
A multipole decomposition of such a projected mass distribution provides 
moments that are identified as power ratios after normalisation by the zeroth moment.
In practice, the power ratio analysis is applied to the surface brightness 
distribution.

The moments $P_m$ are defined as
\begin{equation}
P_0 = \left[ a_0 \ln (R_{\rm ap}) \right]^2
\end{equation}
\begin{equation}
P_m = \frac{1}{2 m^2 R_{\rm ap}^{2m} } \left( a_m^2 + b_m^2 \right)
,\end{equation}
\noindent where $R_{\rm ap}$ is the aperture radius in units of \rfive{}. 
The moments $a_m$ and $b_m$ are calculated by
\begin{equation}
a_m(r) = \int_{r \le R_{\rm ap}} d\vec{x} ~S(\vec{x}) ~r^m \cos (m\phi)
\end{equation}
\noindent and
\begin{equation}
b_m(r) = \int_{r \le R_{\rm ap}} d\vec{x} ~S(\vec{x}) ~r^m \sin (m\phi), 
\end{equation}
where $S(\vec{x})$ is the X-ray surface brightness image, and the integral 
extends over all pixels inside the aperture radius. 
Thus, $a_0$ in Eq. (1) is the total radiation intensity inside the 
aperture radius.

Since all $P_m$ are proportional to the total intensity of the X-ray image, 
all moments are normalised by $P_0$ , resulting in the so-called 
power ratios, $P_m/P_0$.
For brevity, we refer to $P_m/P_0$ as $P_m$ in the rest of the paper. 

We used \pt{} as one measure of the substructure degree since it is 
the lowest moment that measures geometric asymmetries above an ellipticity. 
We calculated the uncertainty of the power ratio measurement
and the influence of photon noise with end-to-end Monte Carlo simulations 
in which an additional Poisson noise was imposed on the count images with 
background. 
We interpreted the variance of the power ratio results from the simulations 
as the measurement uncertainty and subtracted the additional noise 
found in the mean of all simulations compared to the observations from the 
observational result.
Further technical discussions are found in~\cite{chon12}.

\subsection{Centre shifts}

The centre shift measures the stability of the X-ray centre calculated at 
different radii and is formulated as~\citep{poole06}
\begin{equation}
w~ =~  \left[ \frac{1}{N-1}~ \sum \left( \Delta_i -  \langle 
\Delta \rangle \right)^2 \right] ^{1/2} ~\times~ \frac{1}{r_{500}}, 
\end{equation}
\noindent where $\Delta_i$ is the distance between the mean centroid and 
the centroid of the $i$th aperture. 

The centroid of each aperture is found by determining the centre of mass 
of the photon distribution within this aperture.
The resulting \cs{} is then the standard deviation of the different centre 
shifts (in units of \rfive{}).
We used the mean centroid value of all apertures as the reference centre.

The uncertainties in the \cs{} parameter were determined with the same 
simulations as those of the power ratios, 
that is, by using Poissonised resampled cluster X-ray images. 
The standard deviation of the \cs{} parameter in the simulation was used as 
an estimate of the measurement uncertainties. 
We did not use the noise-bias-subtracted \cs{} parameter as in the case 
of the power ratios since the bias correction is mostly much smaller than 
the errors and the bias correction does not shift the \cs{} parameter 
to alter the classification of the cluster morphology.

The end-to-end simulations for the power ratios and centre shifts ensures that, 
for example, the systematics introduced by the photon shot noise is properly 
taken into account in the parameter uncertainties.

\subsection{Morphological classification}

\begin{figure}
  \includegraphics[width=\columnwidth]{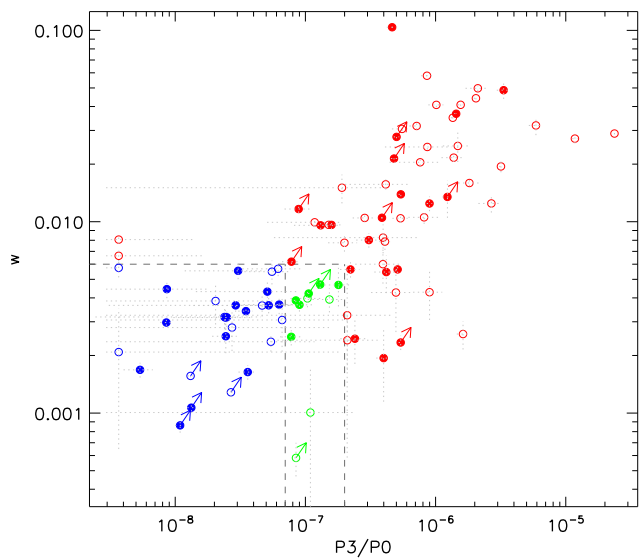}
  \caption{
    Structural parameters, \cs{} vs. \pt{} , for the distributions of the 
    disturbed (red), intermediate (green), and relaxed (blue) clusters.    
    Solid circles are the clusters that also belong to FLS1 above 
    a flux limit of $10^{-11}$~\fxunit{}.
    For the clusters whose \rfive{} is not covered by the XMM-Newton
    observations, the measured values are lower limits, as indicated 
    by upward arrows.
    The three morphological classifications are based on the three dashed
    lines.
  }
  \label{fig:wp3}
\end{figure}

The measured \cs{} and \pt{} values for the \vls{} clusters are shown in 
Fig.~\ref{fig:wp3}.
Both parameters are generally correlated with some scatter.
We note that for the clusters for which \rfive{} is not fully covered 
by the XMM-Newton observations, the structural analysis was only calculated
out to the minimum available fraction of \rfive{} measured from the
centre of the aperture.
This is to ensure that no artificial asymmetry is introduced when
the aperture does not fully cover the cluster. 
Hence in these cases, the measured \cs{} and \pt{} are lower limits 
of the true values since we normalised both parameters by \rfive{} 
instead of their available fraction, and they are represented by upward 
arrows. 

As in~\cite{chon12,chon16}, we defined three boundaries to classify 
the clusters into three distinct categories:
disturbed (\cs{} > 0.006 or \pt{} > 2$\times$10$^{-7}$), 
intermediate (\cs{} < 0.006 and 7$\times$10$^{-8}$ < \pt{} < 2$\times$10$^{-7}$), and
relaxed (\cs{} <  0.006 and \pt{} < 7$\times$10$^{-8}$).
The number of clusters based on this classification is listed in Table 1 under 
the heading VLS, and the same information is given for two flux-limited samples 
that are constructed from \vls{}, namely FLS1 and FLS2.
For comparison, FLS1 clusters are represented by the filled circles in 
Fig.~\ref{fig:wp3}.
There are twice more disturbed clusters than relaxed clusters in \vls,{}
while the two FLSs have more comparable number statistics for 
both populations.
Figure~\ref{fig:wp3} and Table 1 show that the average morphology of 
the clusters in the volume-limited sample is very different to 
those found in FLSs. 

\begin{table}
\begin{center}
\centering
\caption{Morphological classification in our VLS and two FLSs.
FLS1 is constructed with the flux limit of 10$^{-11}$ and FLS2 
with 1.3$\times$10$^{-11}$~\fxunit{}.
The number of clusters in the intermediate class is the difference
between the total and the sum of the disturbed and relaxed clusters.
}
\begin{tabular}{l l l l}
\hline
\multicolumn{1}{l}{Morphology} & 
\multicolumn{1}{l}{VLS} & 
\multicolumn{1}{l}{FLS1} &
\multicolumn{1}{l}{FLS2} \\
\hline
\rule{0pt}{3ex}Disturbed & 56 & 23 & 19   \\
Relaxed                               & 27 & 21 & 18   \\
\hline
\rule{0pt}{3ex}Total          & 93 & 51 & 42 \\
\hline
\end{tabular}
\end{center}
\label{tab:tab1}
\end{table}

\section{Cool-core fraction}

The fraction of cool cores (CC) in an X-ray cluster sample was used to study 
the degree of CC bias and as an indicator for cluster dynamics. 
Because a CC cluster does not necessarily indicate a dynamically relaxed 
state, we simply make a comparison of the fraction of CCs in the flux-
and volume-limited samples.

\begin{figure}
  \includegraphics[width=\columnwidth]{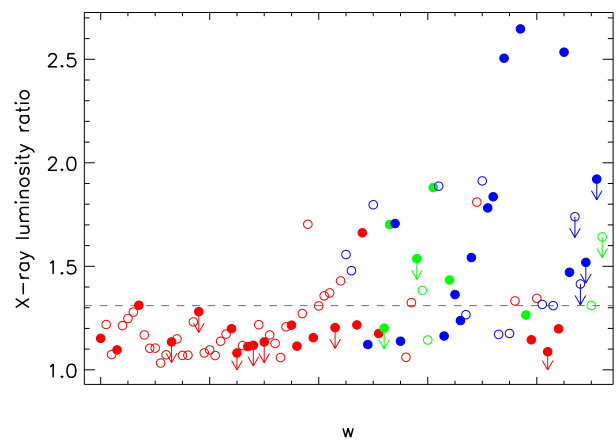}
  \vspace{-0.7cm}
  \caption{
    Luminosity ratio of ReVols clusters (all) in comparison to a FLS1
    (solid) ordered by \cs{} values, which decrease from left to right.
    Colours follow the morphological classification assigned 
    in Fig.~\ref{fig:wp3}.
    For the clusters whose \rfive{} is larger than the XMM-Newton field-of-view
    the measured ratio is an upper limit, as indicated by downward arrows.
    The dashed line is drawn at the value of 1.31, which divides the clusters
    into CC and non-CC.
  }
  \label{fig:ccf}
\end{figure}

The cores of clusters in \vls{}, defined by 0.1$\times$ \rfive{}, are well 
resolved in the XMM-Newton observations.
Hence we calculated the luminosity ratio between the total luminosity 
in the aperture out to the clusters \rfive{} and that measured 
in the same aperture without the core region~\citep{rexcess_sub}. 
Analogous to the substructure calculations, \rfive{} was not fully
covered in the observations for some largest clusters. 
In this case, the surface brightness ratio was calculated again out to the available
aperture and was normalised to \rfive{}.
Since the core was always observed, the luminosity ratio becomes an upper limit, 
and this is indicated by a downward arrow in Fig.~\ref{fig:ccf}. 
We adopted the nominal luminosity ratio of 1.31 as used in~\cite{rexcess_sub} 
to divide the CC population and the non-CCs.
Figure~\ref{fig:ccf} and Table 2 show that there are more non-CC clusters 
in \vls,{} while the FLS clusters (filled circles) have more comparable 
number statistics between the two populations. 

\begin{table}
\begin{center}
\centering
\caption{
Number of clusters in each class.
The samples are identical to those described in Table 1.
}
\begin{tabular}{l l l l}
\hline
\multicolumn{1}{l}{} & 
\multicolumn{1}{l}{VLS} & 
\multicolumn{1}{l}{FLS1} &
\multicolumn{1}{l}{FLS2} \\
\hline
\rule{0pt}{3ex}Non-CC & 57  & 27  & 22 \\
Cool Core                      & 36  & 24  & 20 \\
\hline
\rule{0pt}{3ex}Total      & 93  & 51  & 42 \\
\hline
\end{tabular}
\end{center}
\label{tab:tab2}
\end{table}

\section{Distinct distributions of X-ray luminosity }

Since we expect that CC clusters are on average more luminous than non-CC
clusters for a given cluster mass, we studied the distributions of the X-ray
luminosity in more detail for our sample.

\begin{figure}
  \includegraphics[width=\columnwidth]{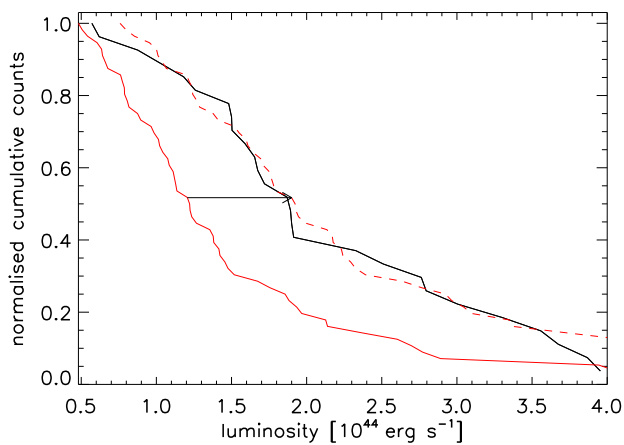}
  \vspace{-0.5cm}
  \caption{
    Normalised cumulative luminosity functions for the disturbed (red) 
    and relaxed (black) clusters.
    The red dashed line represents the distribution for the disturbed
    clusters scaled by 1.57.
  }
  \label{fig:lxdist}
\end{figure}

Figure~\ref{fig:lxdist} shows two normalised cumulative X-ray 
luminosity distributions for the disturbed clusters as a red line 
and for the relaxed clusters as a black line.
The two populations have distinctly different luminosity distributions,
from which we can deduce that there are relatively more luminous clusters
in the sample of relaxed clusters than in the disturbed cluster
sample.
A Kolmogorov-Smirnoff (KS) test shows that it is highly unlikely
that both distributions are the same, with a probability of 0.0003.
This difference in the luminosity distribution can have several reasons.
When we assume that the mass distribution is not drastically different
for the two populations (as found, for example, for the REXCESS sample
and in simulations~\citep{rexcess_sub}), we could attribute this effect
mostly to the luminosity difference for a given mass.
In this case, we expect that the two distributions show a similar shape
with a constant displacement.
Applying a chi-square fit yields a displacement factor of 1.57, as shown by  
the red dashed line in Fig.~\ref{fig:lxdist}.
To avoid edge effects, we fitted in the luminosity range between
0.7 and 3.7$\times$10$^{44}$~erg~s$^{-1}$.
The KS test also finds a best-fit value of 1.59 that maximises 
the probability, that is, the value that minimises the difference between 
the two distributions.
These results imply that the difference in the luminosity distribution
can be explained if relaxed clusters are typically approximately 60\%
more luminous than the disturbed clusters.

An approximate factor of 1.6 can in fact be derived from the 
scaling relations that we presented in Table 2 of~\cite{chon12}.
We fitted scaling relations by dividing the clusters according to 
their morphological classification, finding that there are differences
of 40\% and 20\% in the amplitude of the $L_X$--$T$ and 
in the $M_X$--$T$ relations, respectively, for a fixed slope.
Since the two differences add, this implies that overall we expect a difference of about
60\% in the luminosity between the two populations
for a given mass.
It is likely that this is not the only effect that contributes
to what we observe in \vls,{} and in a forthcoming paper we will 
present scaling relations, which will provide a more complete
picture.

\section{Summary and discussions}

We used two measures of substructures, centre shifts and the third moment 
of the power ratios, to diagnose the degree of substructures in X-ray 
clusters for our ReVols and two flux-limited samples derived from
REFLEX clusters.
As far as we are aware, we present the first numerical evidence that 
clusters in a volume-limited sample are different from a flux-limited 
sample in morphology and in the cool-core fraction with relatively large 
number statistics. 
We find twice more disturbed than relaxed clusters, and CCs do not
dominate \vls{}. 
Thus we do not find that the VLS has a cool-core bias in comparison
to cluster populations from SZE surveys.
The so-called cool-core bias is therefore found in X-ray
cluster samples that are compiled in a flux-limited way.

In the application of X-ray flux-limited cluster samples for cosmological
studies, the use of the scaling relation between X-ray luminosity 
and mass is an important ingredient, and the correction for Malmquist bias 
related to the scatter in the relation is a prerequisite.
In our previous study, we showed that clusters with different morphologies
or dynamical states easily influence scaling relations, as demonstrated 
in Figs. 11 and 12 of~\cite{chon12}. 
In the standard correction for Malmquist bias, the overall
scatter of the $L_X$--M relation is taken into account regardless of morphological
types, and the scatter is typically assumed to have a Gaussian distribution.
Our findings suggest a different way of Malmquist-bias correction that 
may lead to a higher precision in the results because the distribution
of the scatter for a mixed morphological population does not appear
to be Gaussian.
Therefore a procedure where the relation and scatter are determined
independently for disturbed and relaxed clusters while taking
the morphological fractions also into account may provide a more precise
Malmquist-bias correction.
As an alternative to this approach, the excision of the cool-core
region has been suggested as early as~\citet{markevitch98}.
For surveys where the core of clusters cannot be resolved, however, such as 
the RASS, our suggested refinement of bias correction should provide an 
improvement.
Moreover, in the future, the eROSITA survey does not have the imaging
resolution for the clusters at medium and larger distances 
to precisely excise the core, and the suggested procedure may help 
to improve the efficiency of the cosmological test.

\begin{acknowledgements}
  Our research is based on the XMM-Newton facility and data archive
  operated by ESA. 
  We thank for the support of the XMM-Newton team, 
  especially Norbert Schartel, Ignacio de la Calle, and Maria Santos-Lleo.
  HB and GC acknowledge support from the DFG Transregio Program TR33
  and the Munich Excellence Cluster 
  ''Structure and Evolution of the Universe''.  
  GC acknowledges support by the Deutsches Zentrum f\"ur Luft- und Raumfahrt 
  under grant no. 50 OR 1601.
\end{acknowledgements}

\footnotesize{
  \bibliographystyle{aa}
  \bibliography{subs} 
}

\end{document}